\documentclass{appolb}
\usepackage{graphicx}
\usepackage{subfigure}
\usepackage{gensymb}
\begin{document}

\title{
Commissioning of the Inner Tracker of the KLOE-2 experiment
}
\author{A. Di Cicco~\footnote{Roma Tre University, Rome}, G. Morello~\footnote{LNF - INFN, Frascati}
\address{On behalf of the KLOE-2 Collaboration}}

\maketitle
\vspace{-5mm}
\begin{abstract}
The KLOE-2 experiment is undergoing commissioning at the DA$\Phi$NE $e^{+}e^{-}$ collider of the Fra\-sca\-ti National Laboratory of the INFN, after the integration of new detectors in the former KLOE apparatus. The Inner Tracker, a very light detector (material budget $<2\%X_{0}$), is one of the new subdetectors and it is composed of 4 cylindrical triple-GEM with a stereo X-V strips/pads readout.  
The commissioning phase of the Inner Tracker consists in the characte\-rization of the detector response and in its performance evaluation.
The method used to evaluate its detection efficiency is reported, together with some preliminary results.
\end{abstract}
\vspace{-3mm}
\section{Introduction}
After the integration of the new subdetectors~\cite{let, het, kloeIT, CCALT, QCALT} in the KLOE apparatus~\cite{kloe},  the commissioning phase of the KLOE-2 Inner Tracker (IT) has started. This detector is composed of 4 cylindrical triple-GEM coaxial layers~\cite{tdr}, equipped each with a double-view XV-strips/pads readout circuit. The dimension of the active area ($\sim700\times300~\rm{mm^{2}}$) pushed the GEM deliverer (CERN TE-MPE-EM workshop), in collaboration with INFN-LNF groups, to proceed with a new manufacturing technique \cite{alfonsi}, while the assembly procedure has been totally developed at the Frascati National Laboratory (LNF) of the INFN. The Inner Tracker has been installed between the beam pipe and the inner wall of the KLOE Drift Chamber (DC) and will increase the acceptance for low transverse momentum tracks and improve charged vertex reconstruction (especially useful for events with more than two tracks originating from Interaction Point \cite{CPT}). The whole KLOE-2 \cite{kloe-2} apparatus works in a $0.52~\rm{T}$ magnetic field. Although each layer has been tested in a dedicated area at LNF, before their integration in the IT, new tests with the detector in its final position are required. The IT commissioning benefits from the presence of the DC and its excellent tracking performance.
\section{Track reconstruction with the Inner Tracker}
\begin{figure}[h!]
	\begin{minipage}[t]{.48\textwidth}
	\includegraphics[width=1.\textwidth]{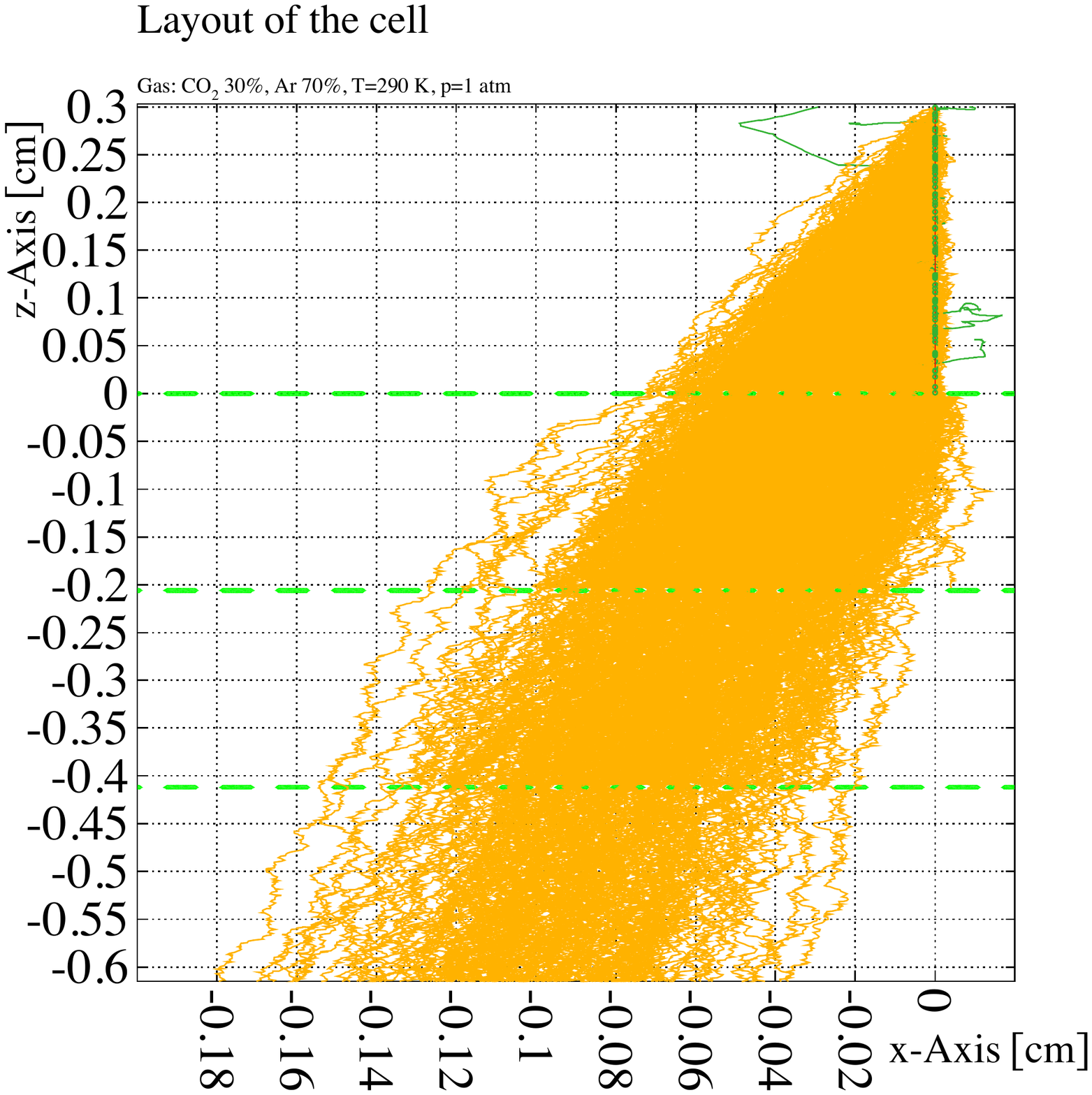}
	\vspace{-18mm}
	\caption{GARFIELD simulation of electrons drift in a $0.52~\rm{T}$ magnetic field with Ar:CO$_{2}$ $70:30$ gas mixture: the track is at the position $x=0$.}
	\label{magfield}
	\end{minipage}
	\hspace{2mm}
	\begin{minipage}[t]{.48\textwidth}
	\centering
	\includegraphics[width=1.05\textwidth]{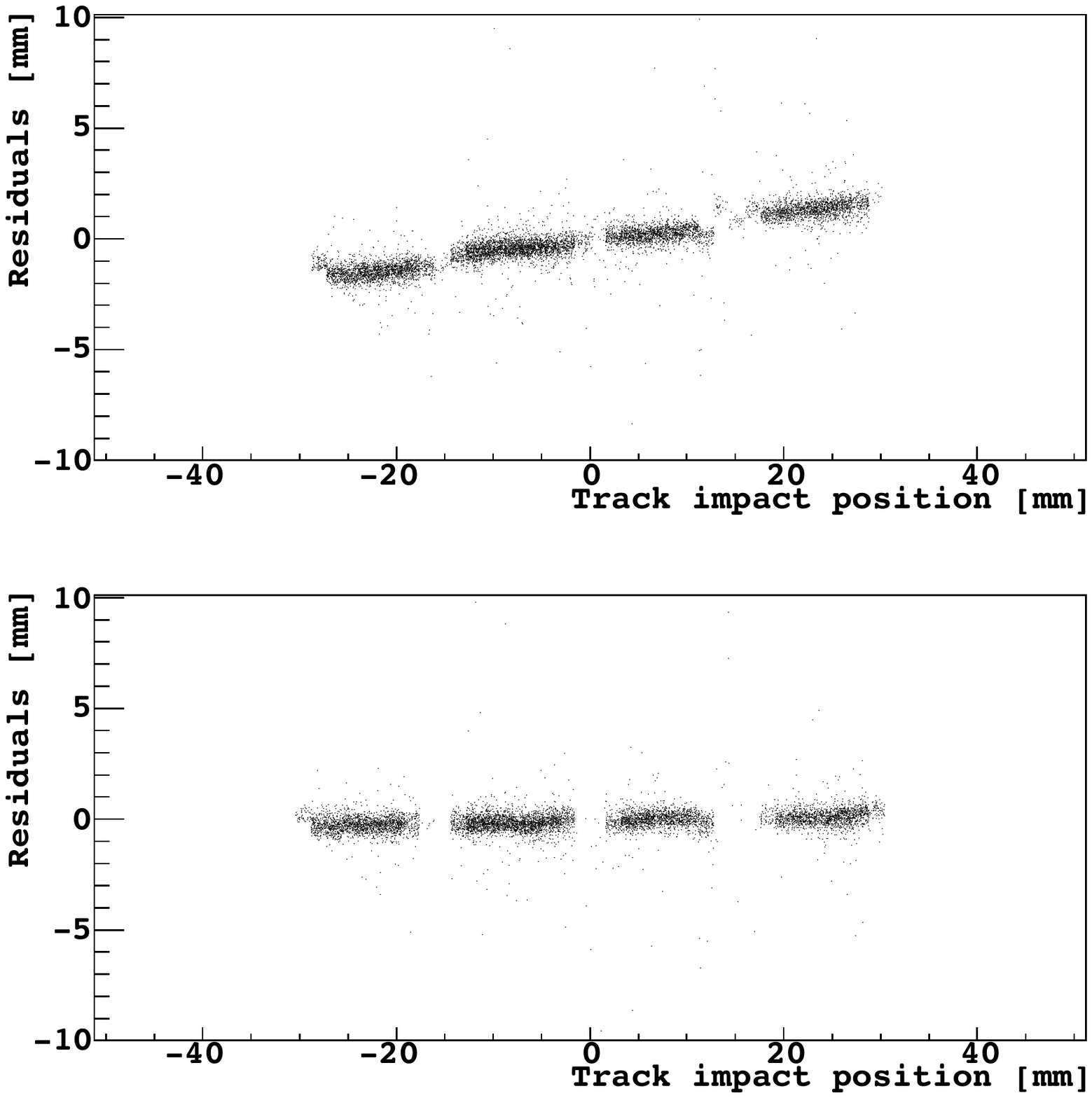}
	\vspace{-18mm}
	\caption{Non-radial tracks effect on the residuals measured with a cylindrical GEM prototype.}
	\label{nonradial}
	\end{minipage}
\end{figure}

Starting from DC hits, tracks are reconstructed including IT hits by using the Kalman filter. Inner Tracker must be aligned and calibrated in order to get the best tracking performance. The presence of the B-field influences the reconstruction since the signal electron cloud experiences the Lorentz force and therefore there is a shift in the position of the fired readout strips with respect to the true position of the intersection between the particle track and the readout plane. Spatial resolution is also affected by non-radial tracks crossing the detector as shown in fig. \ref{nonradial} \cite{tdr}. The combination of the two effects produces a focusing or a defocusing of the electron cloud according to the impact parameter of the track on the cylindrical detectors. These effects must be studied independently: cosmic-ray muon data will be used to evaluate the non-radial correction (without B-field) and the magnetic field influence (with B-field), and Bhabha scattering events will be used to check the effect of the two corrections. The calibration and alignment procedure are presently ongoing. In the next paragraphs the monitoring of the detector status and the efficiency evaluation will be described.
\vspace{-20mm}
\section{Noisy and inefficient strips}
\begin{figure}[th]
	\begin{minipage}{.48\textwidth}
	\centering
	\includegraphics[width=1.\textwidth]{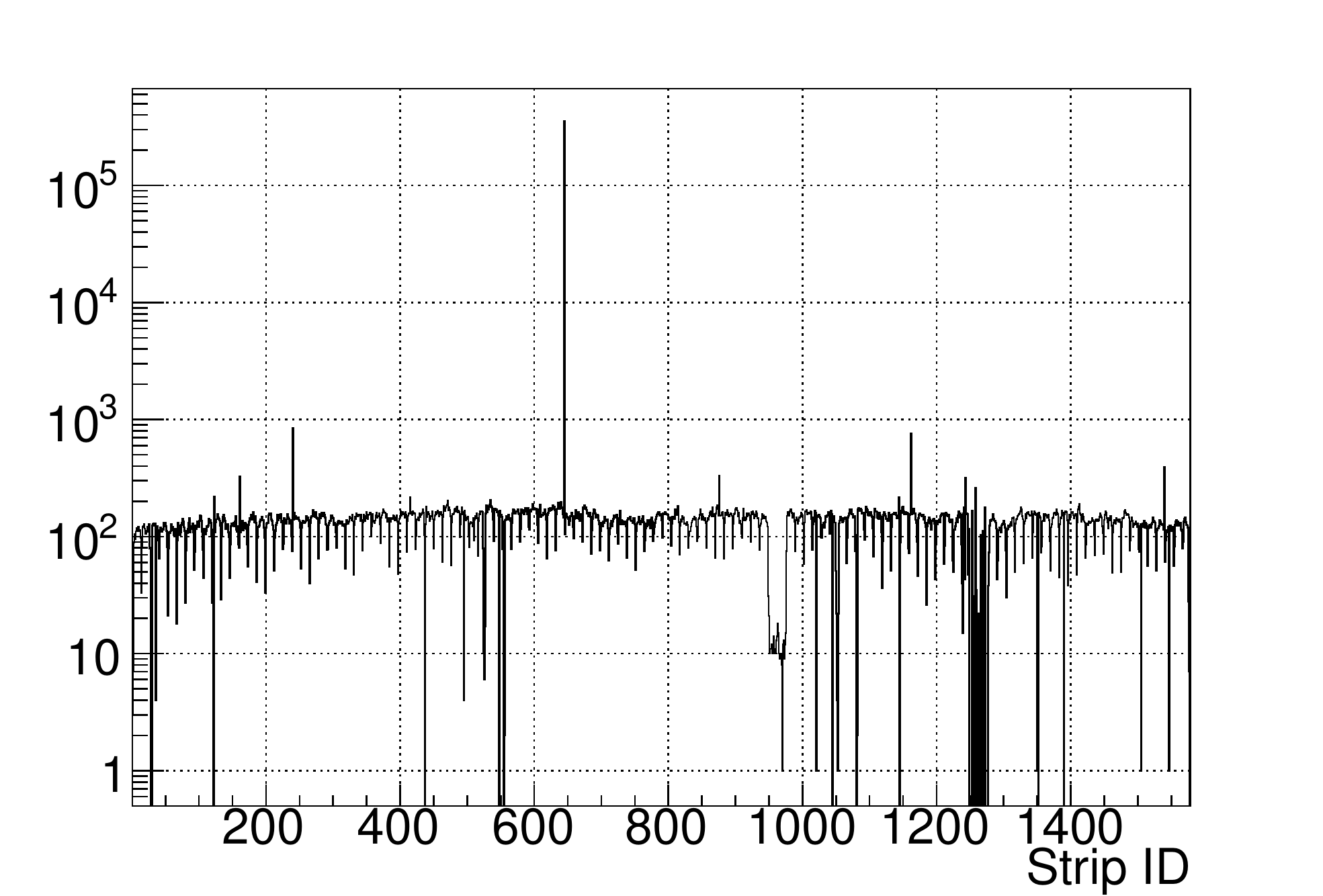}	
	\caption{Layer 2 X-view occupancy distribution.}
	\label{xoccu}
	\end{minipage}	
	\hspace{2mm}
	\begin{minipage}{.48\textwidth}
	\centering
	\includegraphics[width=1.\textwidth]{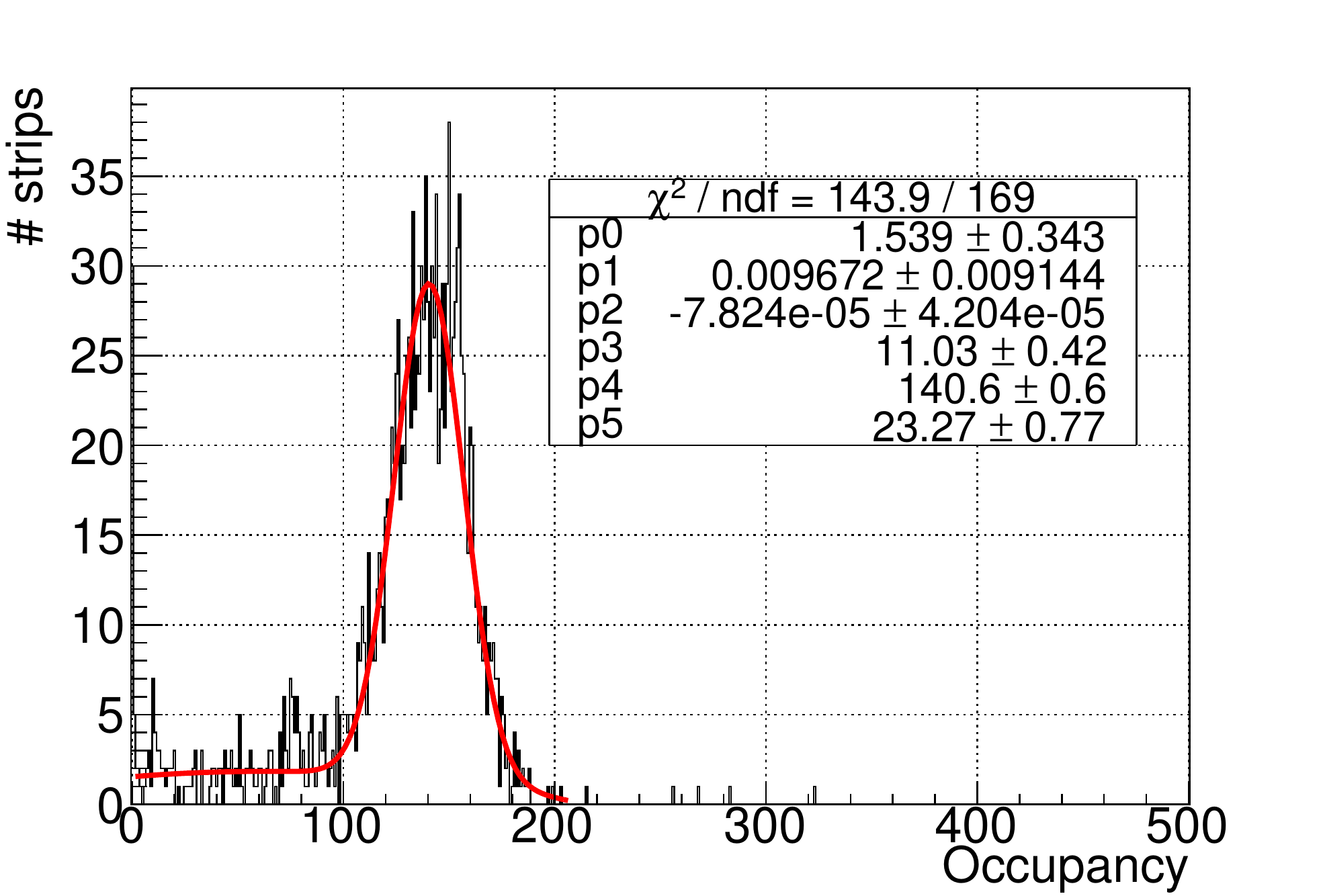}	
	\caption{Layer 2 X-view occupancy profile.}
	\label{xoccuprof}
	\end{minipage}
\end{figure}

The detector status must be monitored in time and taken into account at the analysis stage. The definition of \emph{noisy} or \emph{dead} strips is related to their average occupancy for a single view and for each layer of the Inner Tracker. 
The technique used to evaluate the average occupancy and the definition of the strips status is briefly reported.
\newline
\noindent
\textit{\textbf{X-view.}}
The X-view is given by the axial strips which provide coordinates in the $r$-$\varphi$ plane (transverse plane with respect to the beam line). All these strips have the same length and width (and consequently the same parasitic capacitance) so that,
assuming a uniform exposition to cosmic-ray muons, the profile of the occupancy distribution is expected to be a gaussian function.
In fig.~\ref{xoccu} the occupancy for the X-view strips is shown while its profile distribution is displayed in fig. \ref{xoccuprof}.
This latter is fitted with a combination of a second-degree polynomial (whose parameters are $p_0$, $p_1$ and $p_2$) and a gaussian function.
\vspace{-4mm}
The mean value, $\bar{\xi}$, of the gaussian fit is considered to be the average occupancy of the strips belonging to the layer under study,
while the sigma, $\sigma$, is used to identify the strips status as follows:
let $\xi_{i}$ be the occupancy of the $i$-th strip, then strips with $\xi_{i} = 0$ will be considered \emph{dead}, 
while strips with $\xi_{i}>\bar{\xi}+5\cdot\sigma$ will be considered \emph{noisy}.
The same definitions are applied to V-view strips, although an intermediate step is required as further explained. 
\newline
\noindent
\textit{\textbf{V-view.}}
V-view strips have different lengths and, thus, a non-uniform occupancy distribution.
In order to use the tool previously described it seems natural to normalize the occupancy of the $i$-th strip to its length
and then use the same definitions as for the axial strips.\\
The list of noisy strips is then recored in the database in order to mask them directly at the DAQ initialization stage for data taking and to monitor the time evolution of the noisy and dead strips.
From a reference cosmic-ray run we have $\sim6\%$ of noisy strips and $\sim17\%$ of dead strips considering both views and all the four layers.
\vspace{-3mm}
\section{Inner Tracker Efficiency}
\begin{figure}[t]
	\begin{minipage}{.48\textwidth}
	\centering
	\includegraphics[width=.95\textwidth]{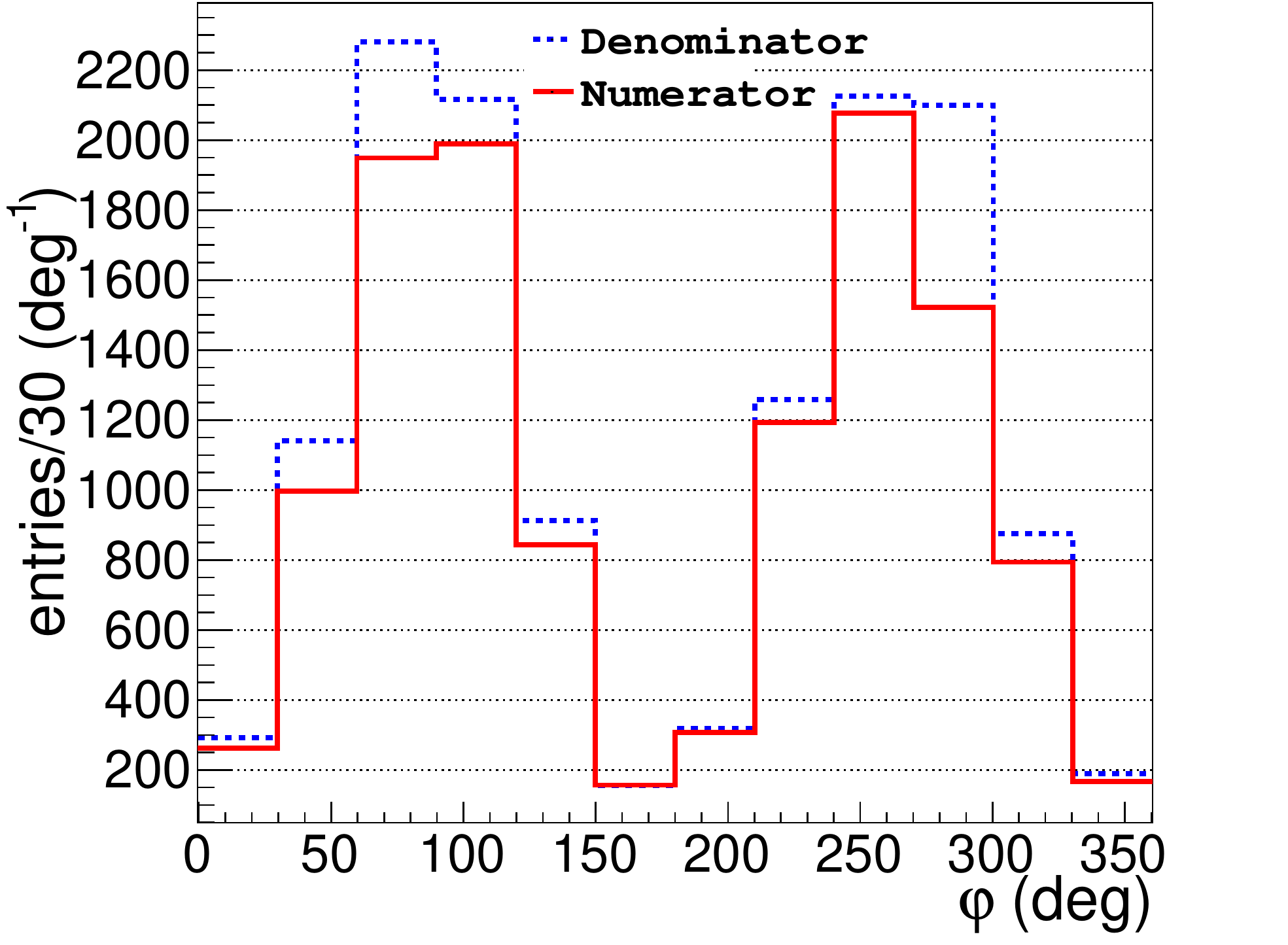}
	\vspace{-1mm}
	\caption{Layer 2 distribution of $\varphi$-position of expected clusters (dashed blue line) and measured clusters (solid red line).}
	\label{fig:fig1}
	\end{minipage}
	\hspace{2mm}
	\begin{minipage}{.48\textwidth}
	\vspace{-9mm}
	\includegraphics[width=1.\textwidth]{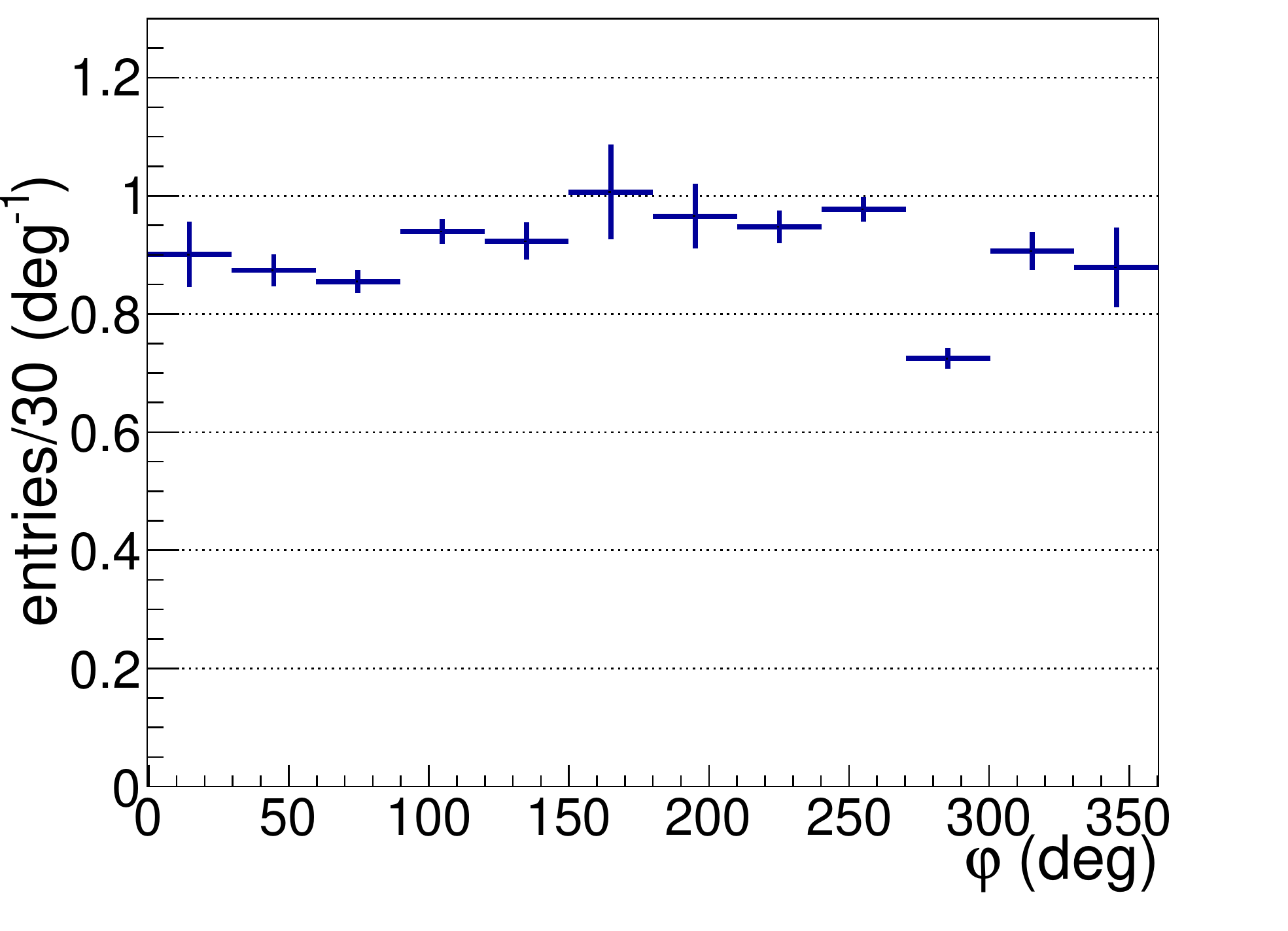}
	\vspace{-7mm}
	\caption{Layer 2 efficiency as a function of the $\varphi$-position in the bending plane.}
	\label{fig:fig2}
	\end{minipage}
\end{figure}

The Inner Tracker efficiency is evaluated by using cosmic-ray muon tracks reconstructed in the Drift Chamber in $0.52~\rm{T}$ magnetic field.
\newline
\noindent
\textit{\textbf{Normalization sample.}}
Tracks reconstructed in the DC are extrapolated to the Inner Tracker (IT) assuming straight lines,
approximation valid if tracks with transverse momentum $p_T > 500\,\rm{MeV}$ are selected. 
Two crossing points with each IT layer and a Point of Closest Approach (PCA) of the track to the beam-line are required, with 
$z_{PCA} <  35\,\rm{cm}$ and $R_{PCA} < 5\,\rm{cm}$, the latter being the PCA radius in the bending plane.
The two crossing points represents the track expected positions on the IT and their angular $\varphi$-position distribution in the bending plane will be the
normalization sample. 
\newline
\noindent
\textit{\textbf{Efficiency evaluation.}}
The efficiency is evaluated as the ratio between the expected and the measured $\varphi$-position distributions.
The latter distributions are obtained by using an algorithm which produces the 3-dimensional coordinates of the detected clusters.
With the position of the reconstructed clusters we can compute the $\varphi$-position in the bending plane.
The numerator of the efficiency is, therefore, the distribution of the measured clusters in the IT (fig.~\ref{fig:fig1}).
A normalization sample of about $6\times10^{6}$ events in which at least a reconstructed track is present has been used to evaluate efficiency. 
Selection criteria reduced the normalization sample to $\sim 3.4\times10^{3}$ events, while $\sim 3.3\times10^{3}$ events enter the numerator.
Layer 2 efficiency as a function of the $\varphi$-position is shown in fig.~\ref{fig:fig2}.
Efficiency values are different from bin to bin, depending on the percentage of dead and noisy strips.
\section{Conclusions}
The commissioning of the Inner Tracker for the KLOE-2 experiment is in progress. The detector operational condition is under optimization and tools to monitor the detector status and measure its efficiency have been developed. Preliminary efficiency measurements have been reported. Calibration and alignment procedures are under development and will use both cosmic-ray muon and Bhabha scattering events.
\section*{\scriptsize{Acknowledgements}}
\scriptsize{
We warmly thank our former KLOE colleagues for the access to the data collected during the KLOE data taking campaign.
We thank the DA$\Phi$NE team for their efforts in maintaining low background running conditions and their collaboration during all data taking. We want to thank our technical staff: 
G.F. Fortugno and F. Sborzacchi for their dedication in ensuring efficient operation of the KLOE computing facilities; 
M. Anelli for his continuous attention to the gas system and detector safety; 
A. Balla, M. Gatta, G. Corradi and G. Papalino for electronics maintenance; 
M. Santoni, G. Paoluzzi and R. Rosellini for general detector support; 
C. Piscitelli for his help during major maintenance periods. 
This work was supported in part by the EU Integrated Infrastructure Initiative Hadron Physics Project under contract number RII3-CT- 2004-506078; by the European Commission under the 7th Framework Programme through the `Research Infrastructures' action of the `Capacities' Programme, Call: FP7-INFRASTRUCTURES-2008-1, Grant Agreement No. 227431; by the Polish National Science Centre through the Grants No. 
DEC-2011/03/N/ST2/02641, 
2011/01/D/ST2/00748,
2011/03/N/ST2/02652,
2013/\-08/\-M/\-ST2/\-00323,
and by the Foundation for Polish Science through the MPD programme and the project HOMING PLUS BIS/2011-4/3. }%
%
%
%
%

\end{document}